\documentclass[submission,copyright,creativecommons]{eptcs}
\providecommand{\event}{TFPIE 2021} 
\usepackage{breakurl}             
\usepackage{underscore}           
\usepackage{listings}
\usepackage{amssymb}
\usepackage{amsmath}
\usepackage{letltxmacro}
\usepackage{verbatim}
\usepackage{xspace}
\usepackage{caption}
\usepackage{subcaption}
\usepackage{amssymb, amsmath}
\usepackage{alltt}
\usepackage{pslatex}
\usepackage{epigraph}
\usepackage{verbatim}
\usepackage{latexsym}
\usepackage{array}
\usepackage{comment}
\usepackage{makeidx}
\usepackage{listings}
\usepackage{indentfirst}
\usepackage{verbatim}
\usepackage{color}
\usepackage{url}
\usepackage{xspace}
\usepackage{stmaryrd}
\usepackage{amsmath, amsthm, amssymb}
\usepackage{graphicx}
\usepackage{euscript}
\usepackage{mathtools}
\usepackage{mathrsfs}
\usepackage{multirow,bigdelim}
\usepackage{subcaption}
\usepackage{placeins}
\usepackage[warn]{textcomp}
\usepackage{tikz}
\usetikzlibrary{arrows,decorations.pathmorphing,backgrounds,fit,positioning,shapes.symbols,chains}
\newcommand{\lama}{$\lambda\kern -.1667em\lower -.5ex\hbox{$a$}\kern -.1000em\lower .2ex\hbox{$\mathcal M$}\kern -.1000em\lower -.5ex\hbox{$a$}$\xspace}

\newcommand*{\SavedLstInline}{}
\LetLtxMacro\SavedLstInline\lstinline
\DeclareRobustCommand*{\lstinline}{%
  \ifmmode
    \let\SavedBGroup\bgroup
    \def\bgroup{%
      \let\bgroup\SavedBGroup
      \hbox\bgroup
    }%
  \fi
  \SavedLstInline
}

\newcommand{\trule}[2]{\dfrac{#1}{#2}}

\newcommand{\withenv}[2]{{#1}\vdash{#2}}

\newcommand{\llang}[1]{\mbox{\lstinline[mathescape=true]|#1|}}

\newcommand{\sembr}[1]{\llbracket{#1}\rrbracket}

\newcommand{\primi}[2]{\mathbf{#1}\;{#2}}

\lstdefinelanguage{lama}{
keywords={ignore, ref, read, write, for, true, false, fun, case, of, esac, let, in, eta, skip, import, public, infix, infixl, infixr, at, before, after, syntax, var, val, if, then, else, elif, fi, do, while, od},
sensitive=true,
commentstyle=\small\itshape\ttfamily,
keywordstyle=\textbf,
identifierstyle=\ttfamily,
basewidth={0.5em,0.5em},
columns=fixed,
mathescape=false,
fontadjust=true,
literate={->}{{$\to$}}3{=>}{{$\Rightarrow$}}3{=>>}{{$\Rightarrow$\hspace{-0.7em}$\Rightarrow$}}3,
morecomment=[s]{(*}{*)},
basicstyle=\normalsize
}

\lstdefinelanguage{plain}{
keywords={},
sensitive=true,
commentstyle=\small\itshape\ttfamily,
keywordstyle=\textbf,
identifierstyle=\ttfamily,
basewidth={0.5em,0.5em},
columns=fixed,
mathescape=false,
fontadjust=true,
literate={->}{{$\to$}}3{=>}{{$\Rightarrow$}}3{=>>}{{$\Rightarrow$\hspace{-0.7em}$\Rightarrow$}}3,
morecomment=[s]{(*}{*)},
basicstyle=\normalsize
}

\title{Reimplementing the Wheel:\\ Teaching Compilers with a Small Self-Contained One}
\author{Daniil Berezun
\institute{St. Petersburg State University \\ St.Petersburg, Russia}
\institute{JetBrains Research}
\email{d.berezun@2009.spbu.ru}
\and
Dmitry Boulytchev
\institute{St. Petersburg State University \\ St.Petersburg, Russia}
\institute{JetBrains Research}
\email{dboulytchev@math.spbu.ru}
}
\def\titlerunning{Reimplementing a Wheel}
\def\authorrunning{D. Berezun, D. Boulytchev}
\begin{document}
\maketitle

\begin{abstract}
  We report on a one-semester compiler construction course based on the idea of implementing a small self-contained
  compiler for a small model language from scratch, not using other compiler construction frameworks. The course
  is built around an evolving family of languages with increasing expressiveness and complexity, which finally is
  crowned by a language with first-class functions, S-expressions, pattern matching, and garbage collection.
  The code generation technique is based on the idea of \emph{symbolic interpreters}, which allows to implement a
  robust albeit not a very efficient native code generator. We give the motivation for the course, describe its
  structure, and report some results of teaching based on students' post-course surveys.  
\end{abstract}

\section{Introduction}

Compilers are probably the most important tools for software engineers; understanding how compilers work is one of the basic
(if not the first) skills for them to master. Thus, to no surprise, compiler design and construction as a separate subject
often is included in the curriculum in many colleges and universities all over the world.

Since the first compilers were implemented in the middle of the 20th century the craft of compiler-making has advanced tremendously.
Modern compilers implement a wide range of optimizations and perform a large number of transformations on the way to the executables,
and their sources total millions of lines of code. This makes a balance between the completeness of covered topics and the robustness
of reference compiler implementation a tough problem. Typically, a course on compiler construction collects a wide range of topics from
lexical and syntax analysis to SSA form construction, register allocation, instruction scheduling, etc. As a rule, instead of a working
hardware some simplified abstract machine (or even a high-level language like \textsc{C} or \textsc{Scheme}) is taken; alternatively,
an infrastructure like LLVM~\cite{llvm,llvm-site}, GCC~\cite{gcc-site}, GraalVM~\cite{graal-site} or WebAssembly~\cite{webassembly-site}
is used as a backend. This, in turn, sweeps a lot of work under the carpet. Even if some real processor is targeted the completeness of
the implemented compiler remains questionable since it requires a lot of testing and debugging to generate a correct code for
``serious'' programs beyond simplistic ones used as smoke tests.

We advocate a different approach aimed at building a complete compiler for a simple, but not a toy language from scratch, not relying on
any compiler construction framework. Moreover, instead of textbook code generation algorithms with advanced instruction selection, register
allocation, and scheduling~\cite{muchnik,morgan} we use a simplified one based on \emph{symbolic interpreters}. Conventional approaches,
while providing ways of generating efficient machine code, are much harder to implement, debug and test, and there is still a matter of
discussion if the students are capable to come up with a robust implementation that can correctly compile large realistic
programs besides the simplistic tests. The approach based on a symbolic interpreter, on the other hand, is
easier to understand, implement, debug and test, and it is not as error-prone as the conventional ones at the price of less efficient code
being generated. Yet it provides a way to implement a full-fledged compiler from the source language to a real working machine code.
We argue that after a simplistic but complete compiler is implemented more advanced methods can be easily mastered later by those who
actually strive to work in the real compiler industry.

The choice of target and implementation languages is the first important decision to be made in the preparation of a compiler course.
In various other areas of science, there is a common practice to devise a specific set of artificial exercises to demonstrate
the application of certain techniques or develop certain skills. Thus, inventing a language that would serve as a toolbox of features and
constructs called to demonstrate the most relevant techniques in compiler construction seems like a natural idea.

For the course we have developed a language called ``\lama'' (pronounced ``\emph{lamah}'')~\cite{lama-site}, which is an acronym for ``$\lambda$-\textsc{Algol}''
since the language inherited the shape of its syntactic constructs from \hbox{\textsc{Algol-68}}. In a nutshell, \lama is an
\textsc{Algol}-like language with first-class functions and automatic memory management in the form of garbage collection; we describe
it in more detail in Section~\ref{sec:lama}. The compiler for \lama was initially written in \textsc{OCaml} for \textsc{x86-32/Linux}
platform relying only on binutilities and internally using \textsc{gcc} as a driver. For several years the students were implementing
their compilers in \textsc{OCaml}; however, during the recent three semesters, we switched the implementation language to \lama itself,
which opens a way for bootstrapping. We consider using the same language both as the source and implementation one as an important
advantage for a number of reasons. First, the ability to be used as implementation language is a strong argument for the maturity of
the language and, more importantly, the maturity of the methods used in its compiler implementation. Thus, by using exactly the same
language as they are implementing and the same compiler as they are writing the students acquire a justification that they are
studying a working technology. Then, to implement a correct compiler the students need to internalize the knowledge
of the source language semantics. Since the implementation language is the same as the source one, this makes the students better
understand the semantics of the language they use. Finally, when the source language coincides with the implementation one
some tasks can be solved in a simpler way due to the identity of their semantics.

\section{The \lama Programming Language}
\label{sec:lama}

\begin{figure}[t]
  \begin{subfigure}[t]{.4\textwidth}
  \begin{lstlisting}[basicstyle=\scriptsize]
import List;
import Fun;

public infixl => before $ (x, f) {
  fun (state) {
    case x (state) of
      [state, x] -> [state, f (x)]
    esac
  }
}

public infix =>> at => (x, f) {
  fun (state) {
    case x (state) of
      [state, x] -> f (x) (state)
    esac
  }
}

public fun returnST (x) {
  fun (state) {[state, x]}
}    
  \end{lstlisting}
  \caption{State monad implementation}
  \label{fig:state-monad}
  \end{subfigure}
  \begin{subfigure}[t]{.6\textwidth}
  \begin{lstlisting}[basicstyle=\scriptsize]
var simpleStmt = memo $ eta syntax (
      kSkip                                  {Skip}                                             
    | x=lident s[":="] e=exp                 {Assn (x, e)}                     
    | kRead   x=inbr[s("("), lident, s(")")] {Read (x)}        
    | kWrite  e=inbr[s("("), exp   , s(")")] {Write (e)}       
    | kWhile  e=exp b=inbr[kDo, stmt, kOd]   {While (e, b)}    
    | kDo     b=stmt kWhile e=exp kOd        {DoWhile (b, e)}  
    | -kIf ifPart -kFi
    ),
    elsePart = memo $ eta syntax (
      empty {Skip}
    | -kElse stmt
    | -kElif ifPart
    ),
    thenPart = memo $ eta syntax (-kThen stmt),
    ifPart   = memo $ eta syntax (
      cond=exp th=thenPart el=elsePart {If (cond, th, el)}
    ),
    stmt     = memo $ eta syntax (
      simpleStmt
    | s1=simpleStmt s[";"] s2=stmt {Seq (s1, s2)}
    );
  \end{lstlisting}
  \caption{An example of syntax description}
  \label{fig:syntax-def}
  \end{subfigure}
  \caption{\lama Samples}
  \label{samples}
\end{figure}

\lama borrows the syntactic shape of operators from \textsc{Algol-68}~\cite{A68}; \textsc{Haskell}~\cite{haskell} and
\textsc{OCaml}~\cite{ocaml} can be mentioned as other languages of inspiration. The general characteristics of the language are:

\begin{itemize}
\item procedural with first-class functions~--- functions can be passed as arguments, placed in data structures,
  returned and ``constructed'' at runtime via closure mechanism;
\item with lexical static scoping;
\item strict~--- all arguments of function application are evaluated before function body;
\item imperative~--- variables can be re-assigned, function calls can have side effects;
\item untyped~--- no static type checking is performed;
\item with S-expressions and pattern-matching;
\item with user-defined infix operators, including those defined in local scopes;
\item with automatic memory management (garbage collection).
\end{itemize}

The main purpose of \lama design is to present a repertoire of constructs with certain runtime behavior and
relevant implementation techniques. The lack of a type system (a vital feature for a real-world language
for software engineering) is an intentional decision that allows demonstrating an unchained diversity
of runtime behaviors, including those which a typical type system is called to prevent. On the other hand,
the language can be used in the future as a raw substrate to apply various ways of software verification (including
type systems).

In addition to a conventional set of constructs, \lama incorporates an extension to embed syntax definitions
in the form of semantic-extended EBNF into the programs. These definitions are converted into the compositions
of parser combinator applications from \lama standard library. In Fig.~\ref{samples} two \lama samples
are given: a definition of state monad from the standard library~(Fig.~\ref{fig:state-monad}) and an excerpt from
\lama parser written in \lama itself~(Fig.~\ref{fig:syntax-def}).

\begin{figure}[t]
  \centering
  \scalebox{1.5}{  
\tikzstyle{arrow} = [thick,->,>=stealth]  
\begin{tikzpicture}
  [node distance = 2cm, auto,font=\footnotesize,
    every node/.style={node distance=1cm},
    comment/.style={rectangle, inner sep= 5pt, text width=4cm, node distance=0.25cm, font=\scriptsize\sffamily},
    force/.style={rectangle, draw, fill=black!10, inner sep=5pt, font=\bfseries\scriptsize\sffamily}]
  \node (source) [force]                    {\scalebox{0.8}{Source Code}};
  \node (AST)    [force, right = of source] {\scalebox{0.8}{AST}};
  \node (SM)     [force, right = of AST]    {\scalebox{0.8}{SM Code}};
  \node (x86)    [force, right = of SM ]    {\scalebox{0.8}{Native Code}};
  \node (ASTint) [font=\tiny, below = of AST]    {Source Interpreter};
  \node (SMint)  [font=\tiny, below = of SM]     {SM Interpreter};
  \node (x86h)   [font=\tiny, below = of x86, rounded corners, rectangle, draw, dashed, text width=1cm, text centered]    {x86-32};
  \draw [arrow] (source)       -- node [font=\tiny, text width = 0.8cm] {Parser}          (AST);
  \draw [arrow] (AST)          -- node [font=\tiny, text width = 0.8cm] {SM Compiler}     (SM);
  \draw [arrow] (SM)           -- node [font=\tiny, text width = 0.8cm] {Native Compiler} (x86);
  \draw [arrow] (AST.south)    -- (ASTint.north);
  \draw [arrow] (SM.south)     -- (SMint.north);            
  \draw [arrow] (x86.south)    -- (x86h.north);          
  \end{tikzpicture}}
  \caption{The structure of reference \lama compiler}
  \label{fig:structure}
\end{figure}

\begin{figure}[t]
  \begin{subfigure}[t]{\textwidth}
    \begin{lstlisting}
                            printf ("Hello, world!\n")
    \end{lstlisting}
    \caption{Source code}
    \label{fig:source}
  \end{subfigure}
  \begin{subfigure}[t]{0.5\textwidth}
    \begin{lstlisting}[language=plain,basicstyle=\scriptsize]
LABEL ("main")
BEGIN ("main", 2, 0, [], [], [])
STRING ("Hello, world!\\n")
CALL ("Lprintf", 1, false)
END      
    \end{lstlisting}
    \caption{Stack machine code}
    \label{fig:stack-machine}
  \end{subfigure}
    \begin{subfigure}[t]{0.5\textwidth}
    \begin{lstlisting}[language=plain,basicstyle=\scriptsize]
        .globl	main
        .data
string_0:.string "Hello, world!\n"
main:
# BEGIN ("main", 2, 0, [], [], []) / 
# STRING ("Hello, world!\\n") / 
	movl	$string_0,	%ebx
	pushl	%ebx
	call	Bstring
	addl	$4,	%esp
	movl	%eax,	%ebx
# CALL ("Lprintf", 1, false) / 
	pushl	%ebx
	call	Lprintf
	addl	$4,	%esp
	movl	%eax,	%ebx
# END / 
	movl	%ebx,	%eax
Lmain_epilogue:
	movl	%ebp,	%esp
	popl	%ebp
	xorl	%eax,	%eax
	ret
    \end{lstlisting}
    \caption{Native code}
    \label{fig:native-code}
  \end{subfigure}
  \caption{An example of a program, stack machine code and native code}
  \label{fig:compilation-example}
\end{figure}


\section{The Structure of the Compiler}
\label{sec:compiler}

The current implementation of \lama contains a native code compiler for \textsc{x86-32}, written in \textsc{OCaml} ($\approx 3000$ LOC),
a runtime library with garbage-collection support, written in \textsc{C} ($\approx 1000$ LOC), and a small
standard library, written in \lama itself ($\approx 900$ LOC). The native code compiler uses \textsc{gcc} as a driver.
The standard library implements a minimalistic set of features needed to fulfill all the assignments for the course:
a set of collections, implemented as AVL trees, list- and array-processing functions, basic file operations, functional
programming primitives for lazy evaluation, function application/composition/fixpointing, etc., and the implementation
of monadic CPS parser combinators with memoization~\cite{johnson,meerkat}, which support left recursion and are
capable of recognizing all context-free languages.

The structure of the compiler is shown in Fig.~\ref{fig:structure}. Overall, it maintains the generic scheme of compiler
implementation as a sequence of passes each of which performs a transformation of some intermediate representation of
a program being compiled. In our case, there are two such representations: an abstract syntax tree (AST) and
a code for an abstract stack machine (SM). The compilation from AST to stack machine requires two passes (counting
closure conversion); the transformation from SM code into the native one requires one pass. In Fig.~\ref{fig:compilation-example}
an example of ``Hello, world!'' program compilation is presented; besides the source program itself, its SM representation is
shown as well as the native code for \textsc{X86-32}, compiled directly from that representation.

Besides conventional components, the compiler contains two extra ones: a source-level reference interpreter, which literally
encodes the operational semantics, and an interpreter for the stack machine. Thus, a \lama program can be run in three modes: being
interpreted in direct correspondence with operational semantics, compiled to the stack machine code, and compiled to
native code. It is expected that the results of execution in all three modes should coincide for any program.

The decision to include two interpreters in the compiler serves didactic purposes. As student assignments repeat the
implementation of the reference compiler, they would involve implementing these interpreters as well, which serves the
purpose of better understanding of how operational semantics works. In addition, the capability of running a program in different
representations would make it possible to discover and fix errors at earlier stages.

\section{The Structure of the Course}
\label{sec:course}

\begin{figure}[t]
  \centering
  \begin{tabular}{lcl}
    \multicolumn{1}{c}{\textbf{Language}} & \textbf{\textnumero} & \multicolumn{1}{c}{\textbf{Assignments}} \\
    \hline
    \hline
    \multirow{3}{*}{Straight-line Code with Assignments}          &  1 & \textsc{Int}, \textsc{SM}\\
                                                                  &  2 & \textsc{X86}\\
                                                                  &  3 & \textsc{Parser}\\
    \hline
    \multirow{2}{*}{Structural Control Flow}                      &  4 & \textsc{Parser}, \textsc{Int} \\   
                                                                  &  5 & \textsc{SM}, \textsc{X86} \\
    \hline
    Control Flow Expressions                                      &  6 & \textsc{Parser}, \textsc{Int}, \textsc{SM}, \textsc{X86}\\
    \hline
    \multirow{2}{*}{Functions and Declaration Scopes}             &  7 & \textsc{Parser}, \textsc{Int} \\
                                                                  &  8 & \textsc{SM}, \textsc{X86}\\
    \hline
    Arrays and Strings                                            &  9 & \textsc{Parser}, \textsc{Int}, \textsc{SM}, \textsc{X86} \\
    \hline
    Fixnum Arithmetic                                             & 10 & \textsc{X86} \\
    \hline
    S-expressions                                                 & 11 & \textsc{Parser}, \textsc{Int}, \textsc{SM}, \textsc{X86} \\
    \hline
    \multirow{2}{*}{Pattern-matching}                             & 12 & \textsc{Parser}, \textsc{Int} \\
                                                                  & 13 & \textsc{SM}, \textsc{X86} \\
    \hline
    \multirow{2}{*}{Closure Conversion and First-Class Functions} & 14 & \textsc{Parser}, \textsc{Int} \\
                                                                  & 15 & \textsc{SM}, \textsc{X86} \\
    \hline
    Memory Management                                             & 16 & \textsc{Runtime}\\
    \hline
    \hline
    \multicolumn{3}{l}{\textsc{Parser}: parser}\\
    \multicolumn{3}{l}{\textsc{Int}: reference interpreter}\\
    \multicolumn{3}{l}{\textsc{SM}: stack machine interpreter and stack machine compiler}\\
    \multicolumn{3}{l}{\textsc{X86}: native-code compiler}\\
    \multicolumn{3}{l}{\textsc{Runtime}: runtime support library}\\
    \hline    
  \end{tabular}
  \caption{The structure of the course}
  \label{fig:course-structure}
\end{figure}

The course revolves around a set of 12-16 assignments (depending on the actual schedule; it is assumed that an assignment has to be completed within a week).
The assignments are \emph{vertically}-oriented: a certain language feature has to be implemented in all components of the compiler from the parser to
the code generator. Thus, instead of implementing, say, a parser for the whole language first, then a reference interpreter for the
whole language, then stack machine compiler, etc., we ask students to implement in a top-down manner a set of compilers for
an evolving family of languages. We argue that this structure of the course helps the students to internalize relevant compiler
implementation techniques by completing similar sets of tasks with increasing complexity levels over and over again.

As we said earlier, the compiler consists of a reference source-level interpreter, a stack
machine compiler and a stack machine interpreter, and a native code compiler. As a rule, each language in the family is implemented
in two steps: first, a parser and a reference interpreter are implemented, and then the stack machine compiler and interpreter and
native code compiler are added. For each language, its operational semantics is given as well as operational semantics for the current version of
the stack machine (the stack machine evolves as well), which makes it possible to assess the correctness of the
compiler formally (although we do not require this to be done). For all students, this is the first time when they get acquainted with
operational semantics, so we introduce the topic thoroughly.

We supply the students with a set of regression tests, some of which are hand-written and some~---- autogenerated. The set of tests evolves
together with the language to reflect its relevant properties. It is required that the student implementation passes all the tests, and it is
forbidden for students to make changes in the infrastructural parts of the project (e.g. in the Makefiles).

In addition to the tests, we provide the students with the ready-to-use implementation of the infrastructure parts of the compiler. Such components as the driver
(which controls the order of transformations, reads and writes files, parses command-line options, etc.), implementations of symbol tables/environments, the
interfaces between passes, etc., are all important parts of the compiler that define its architecture and are easy to mess with. At the same time,
the implementation of these components has only a distant relevance to the essence of compilation. By pre-supplying these components to the
students we, first, free them from the burden of implementing and debugging the most ``boring'' parts of the compiler; at the same time, we facilitate the
use of the best practices in compiler writing since we provide students with an architecturally solid environment. Finally, using the same
infrastructural parts of the compiler makes it easier for students' implementations to pass the same set of tests since there is no difference in
their interfaces.

The majority of assignments are incremental, meaning, that they amount to adding some functionality to previously completed assignments. There are
some cases, however, when the architecture of the whole compiler changes drastically. We specifically point out these cases and help the
students to go through the refactoring by providing them with code samples that illustrate the transition.

Some tasks might require more than one week to implement, debug and test in full. In this case, we decompose the assignments in such a way that the
hard one comes first, being followed by relatively simpler. We incrementally add more complex tests for the first one in each following assignment and
warn the students, that they will most likely encounter more errors in already implemented parts of the compiler. Thus, we help the students to amortize
their debugging and testing efforts by gradually increasing the complexity and coverage of the tests.

The summary of the course is shown in Fig.~\ref{fig:course-structure}. In the next subsections, we address specifically the concrete languages in the family and describe
the assignments in more detail.

\begin{figure}[t]
  \centering
  \begin{subfigure}{0.6\textwidth}
  \begin{lstlisting}[basicstyle=\scriptsize]
   -- Redefinition of standard infix operators
   infix +  at +  (l, r) {Binop ("+",  opnd (l), opnd (r))}
   infix -  at -  (l, r) {Binop ("-",  opnd (l), opnd (r))}
   infix *  at *  (l, r) {Binop ("*",  opnd (l), opnd (r))}
   infix /  at /  (l, r) {Binop ("/",  opnd (l), opnd (r))}
   infix %  at %  (l, r) {Binop ("%",  opnd (l), opnd (r))}
   infix == at == (l, r) {Binop ("==", opnd (l), opnd (r))}
   infix != at != (l, r) {Binop ("!=", opnd (l), opnd (r))}
   infix <  at <  (l, r) {Binop ("<",  opnd (l), opnd (r))}
   infix <= at <= (l, r) {Binop ("<=", opnd (l), opnd (r))}
   infix >  at >  (l, r) {Binop (">",  opnd (l), opnd (r))}
   infix >= at >= (l, r) {Binop (">=", opnd (l), opnd (r))}
   infix && at && (l, r) {Binop ("&&", opnd (l), opnd (r))}
   infix !! at !! (l, r) {Binop ("!!", opnd (l), opnd (r))}    
  \end{lstlisting}
  \caption{A fragment of deep embedding implementation}
  \label{fig:emb}
  \end{subfigure}
  \hfill
  \begin{subfigure}{0.3\textwidth}
  \begin{lstlisting}
  read ("x") >>
  read ("y") >>
  "z" ::= "y" * "y" >>
  write ("x"+"z")      
  \end{lstlisting}
  \caption{A sample test in the form of deep embedding}
  \label{fig:test}
  \end{subfigure}
  \caption{Deep embedding of straight-line programming language in \lama}
  \label{embedding}
\end{figure}

\subsection{Straight-line Programs with Assignments}

The first language in the family contains a set of expressions and simple statements in the form of assignments, sequential composition, empty operator (``\lstinline|skip|''), and
reading/writing primitives. Expressions can contain variables, integer constants, and thirteen basic arithmetic and logic operators; logic operators
work on integer values \emph{a la} \textsc{C}. Note, on this level the language is already equipped with a full set of arithmetic and logic
operators of \lama. No declarative constructs exist in the language at this stage; all variables are treated as global ones, defined implicitly.

The first assignment involves implementing a reference interpreter for the language, based on its big-step operational semantics, a stack machine interpreter, and a compiler from the source language into the stack machine code. It does not, however, contain such a task as implementing a parser. Instead of the
parser, we use a deep embedding of the language into the \lama. This embedding is implemented by redefining the standard binary operators of \lama
to work with abstract syntax trees instead of integer values; additionally, \textsc{C} preprocessor is used in a minimalistic manner (see Fig.~\ref{fig:emb} for an
implementation snippet and Fig.~\ref{fig:test} for a sample regression test in the form of deep embedding). This approach allows students to concentrate immediately
on essential tasks~--- implementation of interpreter and stack machine compiler~--- without distracting them with so far not very important problem of implementing a
parser.

The second assignment concerns implementing a native code generator. As we've said earlier, we utilize the concept of a symbolic interpreter to generate machine
code. We address this approach in more detail in Section~\ref{sec:symbolic-int}. Since in the first assignment the students have already implemented an
interpreter for the stack machine, the problem should not be very challenging. There are, however, a number of subtleties~--- for example, unlike stack
machine and source-level interpreter in machine code a number of declarations for global variables have to be generated. Another issue concerns the implementation
of binary operators~--- while in source-level interpreter and stack machine the correspondence between operators in target and source languages is one-to-one, in
the machine code more elaborated projections have to be used. Finally, in \textsc{X86} registers are not fully symmetric~--- there are some dedicated registers
that should be used to perform certain operations, and these requirements have to be taken into account. For the second assignment, we provide the students with
a minimalistic set of tests that only verify the basic cases. We add more tests in the next assignments thus providing the students more time for debugging.

Finally, the third assignment consists of implementing a parser for the already implemented compiler. Thus, after three assignments (and three weeks) the students
cover all essential tasks in order to implement a native code compiler for a simple imperative language.

\subsection{Structural Control Flow}

The next language in the family introduces control flow constructs: branching and looping. Two assignments are scheduled for this language: for the first, a parser and
reference interpreter have to be implemented, for the second~--- the compiler and interpreter for the stack machine and native code compiler.

With this language, we introduce the students to the notion of a \emph{syntax extension}, or ``desugaring''. On abstract syntax level, we only introduce
a simplistic branching construct

\begin{lstlisting}[mathescape=true]
    if $c$ then $s_1$ else $s_2$ fi
\end{lstlisting}

However, in concrete syntax we add two derived forms: a reduced one with no ``\lstinline|else|'' part and a multiple-branching form

\begin{lstlisting}[mathescape=true]
    if $c_1$ then $s_1$
    elif $c_2$ then $s_2$
    $\dots$
    elif $c_k$ then $s_k$
    else $s_{k+1}$
    fi
\end{lstlisting}

and require to convert these derived forms during the parsing stage into the basic AST using the obvious rules. Similarly, at the abstract syntax level, we introduce a basic looping
construct

\begin{lstlisting}[mathescape=true]
    while $c$ do $s$ od
\end{lstlisting}

while in concrete syntax we add a derived form

\begin{lstlisting}[mathescape=true]
    for $s_1$, $c$, $s_2$ do $s_3$ od
\end{lstlisting}

which should be converted into

\begin{lstlisting}[mathescape=true]
    $s_1$; while $c$ do $s_3$;  $s_2$ od
\end{lstlisting}

Thus, we show how using a simple method a language can be ``pumped'' with a variety of constructs. And here comes a counterexample: we demonstrate, that this
method cannot be used to implement a post-condition looping construct. Indeed, an obvious conversion rule

\begin{lstlisting}[mathescape=true]
    do $s$ while $c$ od $\leadsto$ $s$; while $c$ do $s$ od $(\star)$
\end{lstlisting}

would lead to an \emph{exponential} growth of AST in the case of nested loops. Thus, instead, we ask the students to devise a direct big-step operational
semantics for post-condition loops in such a way the relation $(\star)$ holds, and implement the construct in a direct style.

\begin{figure}[t]
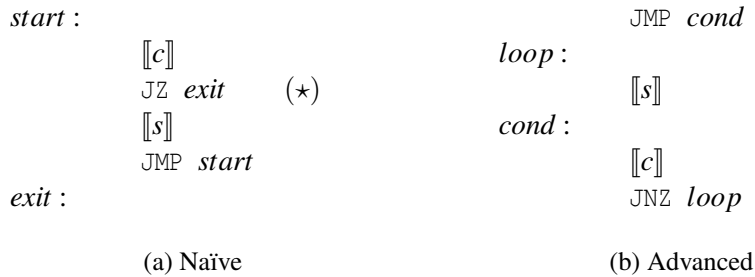

  \begin{subfigure}{0.4\textwidth}
  \begin{lstlisting}[mathescape=true]
    $start:$    
             $\sembr{c}$
             JZ $exit$$\qquad(\star)$
             $\sembr{s}$
             JMP $start$
    $exit:$
  \end{lstlisting}
  \caption{Na\"ive}
  \label{fig:naive-loop}
  \end{subfigure}
  \begin{subfigure}{0.4\textwidth}
  \begin{lstlisting}[mathescape=true]
             JMP $cond$
    $loop:$
             $\sembr{s}$         
    $cond:$    
             $\sembr{c}$
             JNZ $loop$
  \end{lstlisting}
  \caption{Advanced}
  \label{fig:advanced-loop}
  \end{subfigure}
  \caption{A na\"ive and advanced implementations for \lstinline[mathescape=true]|while $\;c\;$ do $\;s\;$ od|}
  \label{fig:naive-advanced}
\end{figure}

In the stack machine, all control flow constructs are represented using labels and conditional/unconditional jumps. The correspondence between stack machine
code and native code is almost one-to-one, so the main job has to be done at the stack machine compilation stage. Here we address two subtleties:

\begin{itemize}
\item A na\"ive conversion of control constructs into the composition of labels and jumps can lead to a situation when an exit from a nested branching/looping
  construct is performed by a chain of jump-to-jump instructions of the length proportional to the nesting level;
\item There is a well-known technique of compiling precondition loops when an additional jump to the condition test is introduced first which allows reducing the number of jumps in the body. In Fig.~\ref{fig:naive-loop} a na\"ive loop implementation is shown with extra jump instruction
  marked by $(\star)$; a better version is given in Fig.~\ref{fig:advanced-loop}. Here $\sembr{\bullet}$ denotes the result of code generation for
  a given construct.
\end{itemize}

For modern processors with branch prediction, neither of these subtleties are essential from the performance standpoint. We, however, still consider
discussing these issues an essential component of the course.

\newcommand{\Refa}{\primi{Ref}{}}
\newcommand{\Val}{\primi{Val}{}}
\newcommand{\Void}{\primi{Void}{}}

\begin{figure}
\renewcommand{\arraystretch}{2}
\renewcommand{\withenv}[2]{{#2}:{#1}}
\[
  \begin{array}{cccl}
    \withenv{\Refa}{\llang{ref $\;x$}} & \withenv{\Val}{x}& \withenv{\Void}{\llang{ignore $\;x$}} & x \in \mathscr X\\
                                      & \withenv{\Val}{z}& \withenv{\Void}{\llang{ignore $\;z$}} & z \in \mathbb N \\
                                      & \trule{\withenv{\Val}{l},\quad\withenv{\Val}{r}} 
                                              {\withenv{\Val}{l\oplus r}} &
                                        \trule{\withenv{\Val}{l},\quad\withenv{\Val}{r}}
                                              {\withenv{\Void}{\llang{ignore $\;l\oplus r$}}}    & \\
                                      &                  & \withenv{\Void}{\llang{skip}}         & \\
                                      & \trule{\withenv{\Refa}{l},\quad\withenv{\Val}{r}}
                                              {\withenv{\Val}{\llang{$l\;$ := $\; r$}}} &
                                        \trule{\withenv{\Refa}{l},\quad\withenv{\Val}{r}}
                                              {\withenv{\Void}{\llang{ignore ($l\;$ := $\; r$})}} & \\
                                      &                  & \withenv{\Void}{\llang{read ($x$)}} & \\
                                      &                  & \trule{\withenv{\Val}{e}}{\withenv{\Void}{\llang{write ($e$)}}} & \\[2mm]
      \trule{\withenv{\Void}{s_1},\quad\withenv{a}{s_2}}{\withenv{a}{s_1;s_2}}&
      \trule{\withenv{\Val}{e},\quad\withenv{a}{s_1},\quad\withenv{a}{s_2}}{\withenv{a}{\llang{if} \;e\; \llang{then} \;s_1\; \llang{else} \;s_2\; \llang{fi}}}&
      \trule{\withenv{\Val}{e},\quad\withenv{\Void}{s}}{\withenv{\Void}{\llang{while $\;e\;$ do $\;s\;$ od}}}&       \\[2mm]
       & \trule{\withenv{\Val}{e},\quad\withenv{\Void}{s}}{\withenv{\Void}{\llang{do $\;s\;$ while $\;e\;$ od}}} & &
  \end{array}
  \]
  \caption{Inference system for expression well-formedness}
  \label{fig:well-formed}  
\end{figure}

\subsection{Control-Flow Expressions}
\label{sec:expressions}

In the previous language, there were two main syntactic categories: expressions and statements. Thus, one could not write

\begin{lstlisting}
    if x then y else 3 fi + z
\end{lstlisting}

or

\begin{lstlisting}
    if x then y else z fi := 3
\end{lstlisting}

etc. As we eventually plan to end up with a language with first-class functions, which is expected to be essentially expression-type, we need to
refactor the language by converting statements into expressions.

The only assignment at this point is the first non-incremental one. Although no new constructs are introduced in the languages, the syntactic
roles of some of them change, which amounts to an essential refactoring of the compiler. Fortunately, this refactoring primarily concerns
parser. All other components of the compiler have to undergo only cosmetic changes.

The main problem which has to be addressed now is the problem of AST well-formedness. Indeed, in the previous language, the well-formedness could easily
be enforced syntactically. Now, however, we need a more elaborated way to prevent one from writing a ``meaningless'' code like

\begin{lstlisting}
    while x do skip od := y
\end{lstlisting}

or

\begin{lstlisting}
    skip + 3
\end{lstlisting}

For this purpose, we equip the language with a simple effect system, which assigns a certain \emph{kind} to each expression. The kinds are propagated in a top-down
manner, and are used both to \emph{check} and \emph{infer} well-formed AST.

There are three kinds: $\Refa$, $\Val$, and $\Void$, which correspond, respectively, to a reference (an expression in assignment position), integer value, or
an empty value. Additionally, there are two specific nodes in the AST~--- ``\lstinline|ignore|'' and ``\lstinline|ref|''~--- which do not have a direct
representation in the concrete syntax. These nodes are \emph{inferred} in order to make the AST well-formed. The topmost kind is always $\Void$.

\begin{figure}[t]
  \begin{subfigure}{\textwidth}
  \begin{lstlisting}[basicstyle=\scriptsize]
var simpleStmt = memo $ eta syntax (
      kSkip                                   {Skip}           
    | x=lident s[":="] e=exp                  {Assn (x, e)}    
    | kRead    x=inbr[s("("), lident, s(")")] {Read (x)}       
    | kWrite   e=inbr[s("("), exp   , s(")")] {Write (e)}      
    | kWhile   e=exp b=inbr[kDo, stmt, kOd]   {While (e, b)}   
    | kDo      s=stmt kWhile e=exp kOd        {DoWhile (s, e)} 
    | ...);
  \end{lstlisting}
  \caption{Simple semantic actions}
  \label{fig:simple-semantics}
  \end{subfigure}
  \begin{subfigure}{\textwidth}
  \begin{lstlisting}[basicstyle=\scriptsize]
var primary  = memo $ eta syntax (
      ...
      loc=pos kSkip                                   {fun (a) {assertVoid (a, Skip, loc)}}                        
    | loc=pos kRead  x=inbr[s ("("), lident, s (")")] {fun (a) {assertVoid (a, Read (x), loc)}}                    
    | loc=pos kWrite e=inbr[s ("("), exp   , s (")")] {fun (a) {assertVoid (a, Write (e (Val)))}}                  
    | loc=pos kWhile e=exp b=inbr[kDo, exp, kOd]      {fun (a) {assertVoid (a, While (e (Val), b (Void)), loc)}}   
    | loc=pos kDo s=exp kWhile e=exp kOd              {fun (a) {assertVoid (a, DoWhile (s (Void), e (Val)), loc)}} 
    | ...);
  \end{lstlisting}
  \caption{Semantics actions in the form of inference system}
  \label{fig:inference-semantics}
  \end{subfigure}
  \caption{Parser implementation with simple semantic actions vs. semantic actions in the form of inference system}
  \label{fig:parser}
\end{figure}

The inference system for kinds is shown in Fig.~\ref{fig:well-formed}; we demonstrate how it works by example. Let us have the following expression:

\begin{lstlisting}
    if x then y else z fi := 2
\end{lstlisting}

The topmost kind is $\Void$, and the topmost construct is an assignment. Thus, the only possible well-formed AST which can be inferred is

\begin{lstlisting}[mathescape=true]
    ignore (if x then y else z fi := 2) : $\Void$
\end{lstlisting}

under the assumptions

\begin{lstlisting}[mathescape=true]
    if x then y else z fi : $\Refa$
\end{lstlisting}

and

\begin{lstlisting}[mathescape=true]
    2 : $\Val$
\end{lstlisting}

The second one checks immediately; the first one is reduced to

\begin{lstlisting}[mathescape=true]
    x : $\Val$
    y : $\Refa$
    z : $\Refa$
\end{lstlisting}

which finally gives us the following well-formed AST:

\begin{lstlisting}
    ignore (if x then (ref y) else (ref z) fi := 2)
\end{lstlisting}

The inference system is implemented directly in the parser. As we noticed earlier, the kinds are propagated in a top-down
manner and the only inference steps are those inserting extra \lstinline|ref|/\lstinline|ignore| AST nodes. This can be
easily implemented by lifting parser semantic actions into kind-accepting functions. Thus, a function that takes a
top-level kind is returned from the parser. By applying this function to $\Void$ we either get a well-formed AST
or fail with an error. An example of parser implementation with simple/lifted semantic actions is shown in
Fig.~\ref{fig:parser}. For the assignment, we give students a partially-refactored parser and ask them to complete
it.

\subsection{Scopes of Definitions and Functions}

The next language in the family adds scopes of definitions and functions. Although functions in this language can be syntactically nested,
at this stage they cannot use the declarations from the enclosing functions yet; later we implement closure conversion, a general
technique for first-class functions.

The declarations come in two flavors~--- for variables (mutable) and values (immutable). Of course, these constructs are introduced at the
expression level in the form of \emph{scope expressions}. Additionally, the existing parser is modified by allowing scope expressions
in a number of contexts (for example, in the branches of conditional expressions, etc.) Two assignments are scheduled for this language. In the first,
scope expressions and functions have to be implemented in the parser and reference interpreter. In the interpreter, the scopes are represented in
a direct way as lists of declarations and their values. In the second assignment, a compiler for stack machine and native-code compiler have to
be implemented; the main work has to be done at stack machine level since the correspondence between stack machine and native code is
almost one-to-one. Unlike reference interpreter, in stack machine all local definitions in a function have to be accumulated and properly
addressed; in addition, in machine code calling conventions have to be respected and activation records of functions have to be
properly organized.

\subsection{Arrays, Strings and Builtins}

\begin{figure}[t]
  \centering
  \includegraphics[scale=0.4]{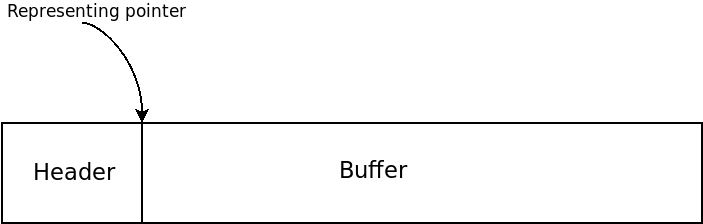}
  \caption{The representation of boxed data}
  \label{fig:boxed}
\end{figure}

All previous languages in the family operated only with scalar integer values. The next one introduces the constructs to deal with
arrays and strings; in particular, it puts to work the notion of \emph{builtin} functions~--- predefined functions which the compiler
is aware of.

Arrays and strings are introduced by adding new kinds of expressions:

\begin{itemize}
\item string, character, and array constants;
\item indexing expressions.
\end{itemize}

It is assumed that indexing expressions can be used both to extract an element from array/string as well as to assign to an element; the kind
inference system described in Section~\ref{sec:expressions} is modified to handle the new kinds of expressions.

As arrays and strings are represented by references, in operational semantics an abstraction of memory is introduced in a conventional
way. It is interesting, however, that in reference interpreter no explicit representation for memory abstractions is needed. Indeed, as
the implementation language coincides with source one, it is already natively equipped with exactly the same abstractions. In other
words, in reference interpreter, we can represent arrays by arrays and strings by strings, which delivers us the expected behavior.

At this stage, for the first time, we need non-trivial support from the runtime library. In all previous languages, the runtime support
library only included two primitives for reading/writing integer values from standard input and output. For arrays and string we need, first,
to define their representation in memory and, second, to provide a number of builtin functions.

The generic layout for arrays/strings is shown in Fig.~\ref{fig:boxed}. Both are represented as a contiguous region of memory (buffer) preceded by
a header. The header contains some supplementary information (in particular, a tag to distinguish strings from arrays, and the length of the buffer;
later a few bits are allocated in the header for garbage collection support). The difference between string and arrays is that for strings
the length of the buffer is calculated in bytes, while for generic arrays~--- in words. In addition, an extra ending zero is kept at the end of
a string. The interesting part is that both strings and arrays are represented in \lama programs by a pointer to the beginning of the buffer, not
to the header. This trick makes it possible to mimic data representation compatible with \textsc{glibc} (in particular, it makes it possible to
pass \lama strings as arguments to \textsc{glibc} functions).
The same layout will be later used for the rest of the data structures in \lama~--- S-expressions and closures.

The set of builtin functions for array/string support includes those for creating arrays and strings, assigning to an element, and taking an
element by an index. All of them are implemented in \textsc{C}, some of them take a variable number of parameters. We provide the students with
these functions implementation. The assignment at this stage includes studying the runtime and implementing arrays/strings at all
levels of the compiler~--- the reference interpreter, extended stack machine, and native code.

\subsection{Fixnum Arithmetic}

In \lama numbers are represented in a fixnum form with the least significant bit of a value always set to 1. This representation makes it possible
to tell pointers apart from scalar values, which is essential for a number of language features support (in particular, pattern matching and
garbage collection).
An assignment to introduce fixnum representation is the second non-incremental one since it requires reimplementation of arithmetic binary
operators support in the native code compiler. No changes to the language itself are made at this stage.

\subsection{S-expressions}

In this assignment, S-expressions are introduced into the language. Similar to arrays/strings, this requires a number of runtime functions
to be implemented. In memory S-expressions are represented similarly to arrays and strings (see Fig.~\ref{fig:boxed}); however, an additional
word is required to store the symbol itself. In the current implementation the first five characters of the symbol are packed into a 32-bit word; thus,
the symbols are distinguished only by their first five characters. 

Since S-expressions are represented similarly to arrays, they can be operated in a similar way. In particular, they can be indexed and their subvalues
can be assigned. For example, the following equalities hold

\begin{lstlisting}
    A (1, 2) [1]         == 2
    A (B (3)) [0][0]     == 3
    length (C (1, 2, 3)) == 3
\end{lstlisting}

\emph{et cetera}.

\subsection{Pattern Matching}

\begin{figure}[t]
  \begin{lstlisting}[basicstyle=\scriptsize]
 var primary  = memo $ eta syntax (
     ...
       loc=pos kSkip {fun (a) {assertVoid (a, returnST $ Skip, loc)}} 
     | loc=pos kWhile e=exp b=inbr[kDo, scopeExpr, kOd] {
         fun (a) {
           assertVoid (a, e (Val)  =>> fun (e) {
                          b (Void) =>  fun (b) {
                            While (e, b)
                          }},
                       loc)
         }
       } 
     | loc=pos kDo s=scopeExpr kWhile e=exp kOd {
         fun (a) {
           assertVoid (a, s (Void) =>> fun (s) {
                          e (Val)  =>  fun (e) {
                            distributeScope (s, fun (s) {DoWhile (s, e)})
                          }},
                       loc)
         }
       }
     | ...);
  \end{lstlisting}
  \caption{Lifting semantic actions into a state monad}
  \label{fig:ST}
\end{figure}

In the support of pattern matching, we again use desugaring in order to simplify the implementation. Namely, we desugar bindings in patterns into
a number of value declarations. For example, the following source-level expression

\begin{lstlisting}
    case f (x + y) of
       A (_, n)     -> n
    |  B (C (k, l)) -> k + l
    esac
\end{lstlisting}

is desugared into

\begin{lstlisting}
    val s = f (x + y);
    case s of
       A (_, _)     -> val n = s[1]; n
    |  B (C (_, _)) -> val k = s[0][0], l = s[0][1]; k + l
    esac
\end{lstlisting}

This approach greatly simplifies the support for pattern bindings by reducing it to already implemented support of nested declaration scopes.
On the other hand, the desugaring has to be implemented properly. As, generally speaking, we need fresh names to bind scrutinees to, we need
to reimplement the parser again, this time lifting the semantic actions into a state monad. The snippet from parser implementation is
shown in Fig.~\ref{fig:ST}.

Two assignments are scheduled for this language. First, the support for pattern matching has to be implemented in the parser and reference interpreter;
this assignment, again, is not incremental since it involves the refactoring of the whole parser. Second, the support for pattern matching in stack machine
and native-code compiler has to be implemented. As a rule, students use a simple top-down branching in their implementations, although we
discuss in the class more elaborated methods for implementation of pattern matching~\cite{Maranget,Maranget1}.

\subsection{Closure Conversion and First-class Functions}

Implementing first-class functions constitutes the final step in the language family evolution. Two assignments are
scheduled for the task. In the first one first-class functions have to be implemented in parser and
reference interpreter. In parser besides the implementation of lambda-expressions call expressions have also to be
generalized to allow arbitrary expressions in the callee position. Surprisingly, the support for first-class functions
in the interpreter is not a hard task at all since we already have the environments collected at the right places.

The second assignment involves implementing closure conversion. This step is performed during the compilation
to the stack machine code. First, a ``draft'' stack machine code is generated with placeholders at closure
initialization and call sites. During this first pass the following supplementary data structures are built:

\begin{itemize}
\item function declarations nesting tree;
\item immediate closure elements (i.e. immediately referenced declarations from enclosing scopes);
\item function reference graph (which function references which).
\end{itemize}

The construction of these data structures is performed within a compilation environment which implementation is
provided as a part of the assignment; the students only need to call certain methods at certain places. After
these data structures are built, the closure conversion can be implemented by propagating immediate closure
elements using a function reference graph and function nesting tree. Then the stack machine code generated
during the first pass is traversed yet again and the placeholders for closure initializations and
calls are replaced with the correct instructions.

Once stack machine code is generated, the implementation of the native-code compiler becomes straightforward. The closures
are represented in a similar way as other data structures, so certain support from the runtime library has to be
provided.

\subsection{Memory Management}

In this assignment, the students are offered to implement a memory manager equipped with one of the basic garbage collection algorithms~--- mark-and-copy.
The assignment consists of two parts. The first one is to implement a two-space heap with a simple sequential allocator which initializes garbage collection
in case the active space is full. The second is a garbage collector implementation consisting of two classical subtasks: root identification and mark-and-copy phase.
In order to identify all roots, it is necessary to traverse the call stack and the static area word-by-word and identify all pointers into the heap.
Due to the fixnum arithmetics, the last bit of each word is used to precisely distinguish pointers from integers. Next, the marking phase is implemented by a
recursive procedure which copies all the live objects into the second space eliminating external fragmentation, lefts a forwarding pointer in the object's old location,
traverses the object for heap pointers, and recursively calls the marking procedure. Finally, the second space has to be traversed in order to change the pointers to new
objects locations. An additional optional challenge is to implement the marking procedure in an iterative manner with support for recovery after the stack overflow.

\section{Code Generation with Symbolic Interpreters}
\label{sec:symbolic-int}

\begin{figure}[t]
  \begin{subfigure}[t]{0.5\textwidth}
    \begin{lstlisting}[basicstyle=\scriptsize]
case i of
  READ       -> case readWorld (w) of
                  [n, w] -> [n : st, s, w]
                 esac
| WRITE      -> case st of
                  n : st-> [st, s, writeWorld (n, w)]
                 esac
| CONST (n)  -> [n : st, s, w]
| LD    (x)  -> [s (x) : st, s, w]
| ST    (x)  -> case st of
                  n : st -> [st, s <- [x, n], w]
                 esac
| ...
esac
    \end{lstlisting}
    \caption{Regular}
    \label{fig:regular-int}
  \end{subfigure}
  \begin{subfigure}[t]{0.5\textwidth}
    \begin{lstlisting}[basicstyle=\scriptsize]
case i of
  READ ->
    case env.allocate of
      [s, env] -> [env, code <+
                         Call ("Lread") <+
                         Mov (eax, s)]
    esac             
| WRITE ->
    case env.pop of
      [s, env] -> [env, code <+
                         Push (s) <+
                         Call ("Lwrite") <+
                         Pop (eax)]
    esac
| CONST (n) ->
    case env.allocate of
      [s, env] -> [env, code <+ Mov (L (n), s)]
    esac
| LD (x) ->
    case env.addGlobal (x).allocate of
      [s, env] -> [env, code <+> move (env.loc (x), s)]
    esac
| ST (x) ->
    case env.addGlobal (x).pop of
      [s, env] -> [env, code <+> move (s, env.loc (x))]
    esac
| ...
esac    
    \end{lstlisting}
    \caption{Symbolic}
    \label{fig:symbolic-int}
  \end{subfigure}
  \caption{Regular vs. symbolic interpreters for stack machine}
  \label{fig:regular-symbolic}
\end{figure}

In this section, we describe the code generation approach which we use throughout the course. As we could see from the previous
sections, the course involves implementing a variety of constructs in a tight schedule. This means that a robust
method for code generation has to be used since otherwise the amount of required debugging and testing efforts would
exceed the students' capacity.

Conventionally, code generation can be logically split into the following subtasks:

\begin{itemize}
\item Instruction selection: a decomposition of source program constructs into a sequence of concrete
  machine instructions.
\item Register allocation: an assignment of concrete registers as operands to selected
  instructions.
\item Instruction scheduling: reordering instructions to make use of intrinsic parallelism of a
  concrete processor.
\end{itemize}

As a rule, these tasks cannot be solved independently: for example, some instructions cannot be chosen due to the lack of
available registers at the moment; the way machine code can be scheduled depends on which concrete instructions were
selected, etc. Thus, in real-world compilers, multiple passes are performed in order to eventually solve all the tasks.
Finally, generating production-quality code involves some combinatorial problems (for example, graph coloring) to be solved,
which, in turn, require specific supplementary data structures to be constructed. All this makes the process of
code generator implementation a very fragile and error-prone task that requires a lot of effort to debug and test
properly. However, in our course the task of implementing a code generator has to be solved multiple times, and
applying conventional approaches would require an unreasonable amount of effort. 

There is, however, a simple method that makes it possible to perform instruction selection and register allocation
in one pass. The approach in question in fact is a part of compiler-writing folklore which is used to be known under
the name ``abstract interpretation'' before the term was taken by a framework in the area of static analysis. The key idea is to use a
symbolic interpreter for a language being compiled which operates on \emph{locations} of data instead of
the data itself. Thus, the task of implementing a code generator reduces to the task of implementing yet another
interpreter, making the whole approach scalable and robust at the price of poorer code quality.

We demonstrate the code generation with the symbolic interpreter at work by the following simple example.
Assume we have a stack machine with the following instructions:

\begin{itemize}
\item \lstinline[mathescape=true, basicstyle=\small]|LD$\;x$|~--- loads a value of a global variable $x$ onto the stack;
\item \lstinline[mathescape=true, basicstyle=\small]|ST$\;x$|~--- stores a value from the top of the stack into a global variable $x$;
\item \lstinline[mathescape=true, basicstyle=\small]|CONST$\;n$|~--- puts a constant $n$ onto the stack;
\item \lstinline[mathescape=true, basicstyle=\small]|BINOP$\otimes$|~--- performs a binary operation ``$\otimes$'' on the top two positions
  of the stack and puts the result back.
\end{itemize}

This stack machine is actually an essential subset of that for the first assignment.
Let us have the following stack machine code:

\begin{lstlisting}[mathescape=true, basicstyle=\small]
    CONST 1
    LD $\;x$
    BINOP +
    ST $\;y$
\end{lstlisting}

A conventional stack machine interpreter would operate on a stack of numbers; the symbolic one, however, operates on a stack
of locations w.r.t. hardware architecture (in our case, \textsc{x86-32}). We can assume that each location is either a
hardware register (\lstinline[basicstyle=\small]|
stack (\lstinline[basicstyle=\small]|S(0), S(1), S(2)| etc., later converted into \lstinline[basicstyle=\small]|-4(
The evaluation steps of the symbolic interpreter update the content of the symbolic stack and emit corresponding
machine code, which can be summarized as the following table:

\vskip5mm
\begin{tabular}{c|c|c|c}
  Stack before                                & Stack machine instruction                               & Stack after                                 & Machine instruction emitted\\  
  \hline
  \lstinline[basicstyle=\small]|{}|           & \lstinline[basicstyle=\small]|CONST 1|                  & \lstinline[basicstyle=\small]|{%eax}|       & \lstinline[basicstyle=\small]|movl $1, 
  \lstinline[basicstyle=\small]|{%eax}|       & \lstinline[mathescape=true,basicstyle=\small]|LD $\;x$| & \lstinline[basicstyle=\small]|{%eax, %ebx}| & \lstinline[basicstyle=\small]|movl $x, 
  \lstinline[basicstyle=\small]|{%eax, %ebx}| & \lstinline[basicstyle=\small]|BINOP +|                  & \lstinline[basicstyle=\small]|{%eax}|       & \lstinline[basicstyle=\small]|addl 
  \lstinline[basicstyle=\small]|{%eax}|       & \lstinline[mathescape=true,basicstyle=\small]|ST $\;y$| & \lstinline[basicstyle=\small]|{}|           & \lstinline[basicstyle=\small]|movl 
\end{tabular}\\[5mm]

The rightmost column accumulates generated code. As one may notice we in fact generated extra instructions in this very short example (in \textsc{x86-32} the effect can
be expressed with a single instruction). However, it's rather clear that with this simplistic method the number of generated machine instructions cannot be less than
the number of stack machine instructions. In Fig.~\ref{fig:regular-symbolic} the snippets from two interpreters of the second assignment~--- regular and symbolic~--- are shown.
In the symbolic one, all the operations on the symbolic stack are implemented by means of an immutable environment (\lstinline|env|); otherwise, the structure of the interpreters is
very similar.

There is a number of considerations that have to be taken into account in order for this method to work properly. First, an essential invariant that has to be
preserved is the order of allocations on the symbolic stack, which has to be fixed. With the fixed order of symbolic stack allocation, we can always recover the contents of the stack from its depth.
A reasonable solution is to allocate the registers first (and in a \emph{fixed} order) and
only when the depth of the stack exceeds the number of available registers we allocate hardware stack slots. The motivation for this is very clear: we use registers
first since they provide better performance. An interesting question is the efficiency of the register allocator, implemented this way. In our case, taking into
account the way the stack machine compiler generates code, registers are assigned in a bottom-up left-to-right traversal of an expression tree. It is known~\cite{arith-trees1,arith-trees2}
that for $n$ registers this method allows generating a machine code with no spilling for a balanced tree with $2^n-1$ nodes, which is not bad at all.

Another issue with the symbolic interpreter is that in fact it sometimes performs steps that are never taken by a conventional one. Indeed, the conventional one
interprets the program in a normal way, making use of available actual data. In particular, it performs conditional and unconditional jumps as prescribed by
their semantics. In contrast, the symbolic interpreter traverses the program once in a top-down manner, thus taking some branches which in fact are never taken by the regular
interpreter~--- for example, while the regular one jumps to an appropriate label when it encounters an unconditional jump, the symbolic one goes to the next instruction. These
observations raise the question if the symbolic interpreter approach can work at all.

Fortunately, it can be shown that stack machine programs generated by the stack machine compiler, in fact, possess the following important invariants:

\begin{itemize}
\item[A:] no unreachable code is introduced;
\item[B:] all instructions are always performed with the same stack depth;
\item[C:] for each label there is at least one preceding instruction that jumps (conditionally or unconditionally) to this label.
\end{itemize}

These properties obviously do not hold for stack programs of general shape; thus, a symbolic interpreter approach works properly only by being coupled with a
specific stack machine compiler. The invariants we mentioned above can be put at work as follows:

\begin{itemize}
\item By invariant A for each label the depth of the stack is the same since labels are instructions.
\item When we encounter a jump (conditional or unconditional) we associate the current stack level with corresponding
  label; by the invariant C, every label will be associated with a certain depth during the top-down traversal prior to visiting.
\item When we encounter an unconditional jump, the next instruction has to be a label (otherwise the next instruction is unreachable, which contradicts the invariant A), and
  this label has already a stack depths associated with it. Thus, we can reconstruct the stack contents and continue.
\item When we encounter a conditional jump, nothing has to be done additionally since the next instruction corresponds to a fallthrough branch.
\end{itemize}

With these invariants preserved there is not much difference in implementation between the symbolic interpreter and the regular one. As the target language
evolves so does the stack machine (new instructions are added gradually), and these instructions have to be interpreted properly by both regular and symbolic interpreters. Since
in each assignment regular interpreter is implemented (and tested) before the symbolic one, writing a code generator becomes rather a routine task.

Another issue that has to be addressed is if this method of code generation worth studying from a didactic standpoint. As we've shown before, by choosing a symbolic interpreter
we traded code quality for the simplicity of the code generation approach; in addition, we completely omitted from consideration the methods which are actually used in production compilers
implementation. The question is if we sacrificed too much. We argue that we actually did not.

From a didactic point of view, while we, indeed, do not consider advanced code generation methods like bottom-up rewriting systems~\cite{burs,burg1,burg} or register allocation
by coloring~\cite{regalloc,regalloc2,regalloc2}, we
still present the students with the tasks of instruction selection and register allocation, albeit in a very simple form. They still need to study the instruction set
of a concrete real-world processor, the assembly language, calling conventions, etc., and they have a certain freedom in controlling the quality of the code they generate. Thus, for the
first-time encounter with the essence of native code generation, they already have enough on their plates. At the same time, while, indeed, the generated code comes 2-3 times slower
than when using more advanced techniques, it actually much faster than could be produced by other methods with comparable simplicity (for example, threaded code).

Finally, we have to note, that actually the stack machine compiler is organized in the same way as a symbolic interpreter of the source language. Thus, we put the same idea to
work twice.

\section{The Course Trivia, Results, and Students' Feedback}
\label{sec:results}

In this section, we present some technical details concerning the organization of the course; we also summarize the results of the post-course anonymous surveys which were collected
during the last three semesters.

The course is being taught at a number of universities in Saint Petersburg, Russia. Initially, those included the Saint-Petersburg University\footnote{\url{https://english.spbu.ru}},
the Saint-Petersburg Department of the Higher School of Economics\footnote{\url{https://spb.hse.ru/en}}, and the ITMO University\footnote{\url{https://en.itmo.ru}}.
However, the unfolding of the coronavirus pandemic forced the course to go online, which made it possible for the students from other cities of Russia to join. For
now, those include Nizhni Novgorod, Novosibirsk, and Moscow.

We put the following soft prerequisites for the course:

\begin{itemize}
\item hardware architecture and assembly language programming;
\item formal languages and grammars;
\item functional programming.
\end{itemize}

We cannot put these as strong prerequisites, however, due to the diversity of curricula in different universities. However, our experience shows that these requirements
are partially fulfilled by the majority of the students.

There are 50-80 attendees each semester. As students come from different universities with different programs and different specializations, and for some of them
the course is mandatory, we offer them a choice of a ``lightning'' division: a test of 100+ questions for the grade C (3 of 5, ``satisfactory'' in Russian grading system),
instead of the regular assignments. Around 1/3 of students usually take this route.

For the rest, we announce a \textsc{github} repository\footnote{The current one: \url{https://github.com/danyaberezun/compilers-2021-spring}}, in which
all the assignments are being published as the course proceeds. Each assignment is put in a separate branch and contains the implementation of the compiler for
relevant language with some parts replaced by placeholders. The students fill in these parts and use provided tests to debug and fix the errors. When they have
all the tests passed locally, they make pull requests to the parent repository. Each pull request is built and test on the \textsc{CI} server via the
\textsc{github} to \textsc{Travis CI} integration. The results of \textsc{CI} build are analyzed and accounted for automatically. Additionally, we selectively
review pull requests in order to identify plagiarism or make recommendations. The deadline for each assignment is set for one week with yet another week added as
a grace period.
In addition to regular weekly lectures, we also provide the students with live support via \textsc{Telegram} chat, where they can ask general and concrete technical questions,
complain about assignment incompleteness, etc.

Our experience shows that among those students who have chosen the ``hard way'' around 25\% form a motivated core, whose members complete the assignments on time, ask questions
in the chat, suggest various improvements, and even implement some optimizations in the compiler besides those required by the assignments. Overall, we do not feel that
the students have any essential problems with the course, and by the end of a semester we, indeed, have 15-20 compilers freshly baked.

By the end of the course, we ask students to provide us some feedback in the form of a survey. The average results over the last three semesters are as follows:

\begin{itemize}
  \item The vast majority qualified the course material as \emph{new} for them (42\%~--- completely new, 58\%~--- mostly new);
  \item 42\% qualified the material as potentially \emph{irrelevant} to their future professional activity; 25\% as relevant, and the
    rest as partially relevant;
  \item An essential fraction complained about the lack of a type system in \lama (prior to the spring of 2020~--- about the type system in \textsc{OCaml});
  \item A self-estimated weekly workload for completing the assignments were estimated as 10+ hours by 20\% of students, 7-9 hours by 20\% of students,
    4-6 hours~--- by the rest 60\%.
\end{itemize}

We would like to complete this section by citing two drastically different students' summaries:

\begin{quotation}
  \textit{Writing a compiler for \lama in \lama was a terrible thing when you had no experience with neither \lama nor its relative language \textsc{OCaml}.}
\end{quotation}

\begin{quotation}
  \textit{A very pleasant thing was that \lama was developed specifically for the course and was truly convenient for compiler implementation, especially if
  one had no prior experience with \textsc{OCaml}.}
\end{quotation}

\section{Related Work}
\label{sec:related}

In this section, we survey some works which we consider related to our course. The amount of existing literature on compiler construction is enormous, and there are hundreds of compiler
construction courses around. Thus, it would be virtually impossible to cite and compare with everyone. Instead, we mention here some of those which we found the most relevant
to the objectives and the structure of our course, or which were served as the source of our inspiration. 

No paper on compiler construction can manage without mentioning the classical ``behemoths'' in the area~\cite{muchnik,morgan,dragon}, including probably the earliest one~\cite{theory}, which, from
the standpoint of modern compiler implementation, now can be only of historical interest. These works were never designed as short (one-semester) introductions
to the subject. Instead, they present a comprehensive and close to complete survey of relevant methods for productive-level compiler construction. An attempt
to implement a compiler using all the covered techniques would end up by another infrastructure like \textsc{GCC} or \textsc{LLVM}. At the same time, these books can be
used to compile a short course by carefully choosing the topics considered relevant. We can categorize our course as a prerequisite for a fearless reading of these books.

A classical book by Wirth~\cite{wirth} and a more recent series of books by Appel~\cite{appel-ml,appel-c,appel-java} take a similar to our approach. In~\cite{wirth} an implementation of
a compiler for \textsc{Oberon-0} language (a simplified \textsc{Oberon}) is considered in details. The compiler itself is written in \textsc{Oberon} and generates an
idealized \textsc{RISC} code, so the approach is very close to ours. In~\cite{appel-ml,appel-c,appel-java}, a compiler from a model language \textsc{Tiger} to \textsc{MIPS} processor is
chosen for the reference. In both approaches, the source languages belong to procedural/object-oriented family (in particular, they do not provide the support for first-class functions,
S-expressions or pattern matching). In addition, in contrast to our approach, the compilers are constructed in a horizontal manner, from complete frontends to complete code generators.

We have also to mention an interesting works~\cite{lcc,lcc1} on \emph{retargetable} compiler for \textsc{ANSI C}. Being written in \textsc{ASNI C} itself, it can be bootstrapped and
represents an interesting attempt to implement a self-contained retargetable compiler. As a code generation engine bottom-up-rewriting system~\cite{burs} is
used, and bin-packing is utilized for register allocation.

In~\cite{incremental} a very similar to ours approach~--- \emph{incremental compiler construction},~--- is reported. A subset of \textsc{Scheme} is implemented
in \textsc{Scheme} as an evolving family of languages. Each language incrementally extends the previous one (hence the name), and the implementation is organized
vertically. However, the code generation follows a more direct approach with no intermediate representation in the form of stack machine code. We argue, that this
representation is important from both pedagogical and technical standpoints as it introduces yet another level of abstraction, allows earlier detection of
certain errors, and makes the retargeting simpler. In addition, no garbage collector is implemented (but this can be considered as rather an incompleteness of
the implementation). An incremental approach has directly influenced the course on compiler construction at Indiana University\footnote{\url{https://iu.instructure.com/courses/1735985}},
and a corresponding textbook is going to be published by MIT Press in 2022.

Finally, we can mention a work on \textsc{ChocoPy}~\cite{ChocoPy}~--- a compiler for a subset of \textsc{Python} to \textsc{RISC-V}, written in \textsc{Python}. The compiler
is implemented in a horizontal manner; in contrast to regular \textsc{Python}, in \textsc{ChocoPy} first-class functions are not supported.

In conclusion, we can state that, to our knowledge, currently there is no course on compiler construction that would incorporate all the following features at the same time:

\begin{itemize}
\item The source language coincides with the implementation one.
\item The compiler is gradually implemented for an evolving family of languages.
\item For each language its operational semantics is provided, as well as for abstract machine used as an intermediate representation.
\item As a result a self-contained code generator for a real hardware processor is implemented with no heavy compiler construction infrastructure used.
\end{itemize}

\section{Conclusion and Future Work}
\label{sec:conclusion}

We shared here our experience on teaching compiler construction with a simple native code compiler implemented from scratch. Several further improvements
to our work can be done. First, we consider extending the native code part to a few new targets, for example, \textsc{x86-64}. Then, we plan to address the performance issues
with the generated code. We consider generating efficient native code with symbolic interpreters an interesting research problem. We can also consider extending the language with
more advanced features~--- objects, continuations, or some means for concurrent/parallel programming.

\bibliographystyle{eptcs}
\bibliography{main}
\end{document}